# Patterned silicon substrates: a common platform for room temperature GaN and ZnO polariton lasers.

J. Zuniga-Perez[1,a], E. Mallet[2], R. Hahe[3], M. J. Rashid[1], S. Bouchoule[4], C. Brimont[3], P. Disseix[2], J. Y. Duboz[1], G. Gommé[1], T. Guillet[3], O. Jamadi[2], X. Lafosse[4], M. Leroux[1], J. Leymarie[2], Feng Li[1], F. Réveret[2] and F. Semond[1]

[1]CRHEA-CNRS, Rue Bernard Gregory, 06560 Valbonne, France
[2]Institut Pascal, PHOTON-N2, Clermont Université, CNRS and Université Blaise Pascal, 24 Avenue des Landais, 63177 Aubière cedex, France
[3]Université de Montpellier 2, CNRS, Laboratoire Charles Coulomb, UMR 5221, 34095 Montpellier, France
[4]LPN-CNRS, Route de Nozay, 91460 Marcoussis, France

A new platform for fabricating polariton lasers operating at room temperature is introduced: nitride-based distributed Bragg reflectors epitaxially grown on patterned silicon substrates. The patterning allows for an enhanced strain relaxation thereby enabling to stack a large number of crack-free AlN/AlGaN pairs and achieve cavity quality factors of several thousands with a large spatial homogeneity. GaN and ZnO active regions are epitaxially grown thereon and the cavities are completed with top dielectric Bragg reflectors. The two structures display strong-coupling and polariton lasing at room temperature and constitute an intermediate step in the way towards integrated polariton devices.

[a] Electronic mail: jzp@crhea.cnrs.fr

GaN and ZnO share many physical properties:[1,2] they crystallize in the wurtzite structure, they are polar materials as well as piezoelectric, they display wide bandgaps in the UV range and compared to other inorganic semiconductors such as GaAs or CdTe, they have large optical oscillator strengths and consequently large exciton binding energies. These two last properties render them very attractive in the context of strong exciton-cavity photon coupling and, in particular, in the context of polariton lasing.[3,4] Indeed, in order to achieve polariton condensation at room temperature at least two conditions must be satisfied: first, the Rabi splitting ($\Omega_{Rabi}$), which is proportional to the square root of the exciton oscillator strength and which determines the depth of the polariton trap in reciprocal space, must be larger than roughly twice the thermal energy at room temperature i.e. larger than about 50meV;[5] and second, polaritons need to have enough time to relax from the excitonic reservoir to the bottom of the lower polariton branch (LPB).[5] This second condition can be fulfilled either by enhancing their relaxation rate, e.g. by increasing the number of polariton-polariton interactions without destroying the strong-coupling regime,[6] or alternatively by increasing their lifetime, which is mainly limited by the cavity photon lifetime (i.e. by the cavity quality factor, Q).

To increase Q, especially in the UV, dielectric distributed Bragg reflectors (DBRs) are recognized to be the best option due to their large refractive index contrast, which enables high absolute reflectivities, large stop-bands and reduced penetration lengths.[7,8,9,10,11] Unfortunately, inserting a high crystalline quality material in between two dielectric amorphous DBRs requires complex processing steps.[7,8,10,11] Instead, DBRs made of semiconductor materials epitaxially grown on a crystalline substrate should enable to maintain a high crystalline quality, but require in turn a much larger number of pairs to obtain Qs similar to those achievable with dielectric materials.[12] The nitrides family (AlInGaN) seems to be more promising for fabricating the DBR than the (ZnMgCdO) one, as it presents a larger refractive index contrast within the family while keeping the same crystalline phase (i.e. wurtzite vs rocksalt).[1,2] However, nitride-based DBRs are not easy to fabricate and face a number of problems. These include: (i) a large built-in strain that accumulates during the DBR growth, due to differences in the lattice parameters of the DBR materials, especially when AlN/GaN or AlN/AlGaN with high Ga content are used;[13] (ii) the formation of cracks due to the thermal expansion coefficient mismatch between nitrides and certain substrates;[14] and (iii) the spatial inhomogeneity of the photonic properties,[12] arising because of thickness or composition fluctuations.

In this letter we propose to grow AlN/AlGaN DBRs on "mesas" as a means to circumvent some of these problems as well as to provide a common platform for the development of GaN and ZnO polariton lasers at room temperature.

With the aim of restricting the nitride DBR growth to predefined areas, a substrate that could be easily patterned needed to be employed. Silicon, which can be easily etched, was the obvious choice. Indeed, the basis of the approach introduced in this letter is the patterning of Si(110) and Si(111) substrates either by wet KOH or dry $SF_6$ etching. Mesas of different sizes, ranging typically from $50 \times 50$ $\mu m^2$ to $500 \times 500$ $\mu m^2$, and of different geometries (triangular, hexagonal, diamond-like, etc.) were initially etched onto the substrates;[15] then, a 30 pair (or 30.5 pair) crack-free $AlN/Al_{0.2}Ga_{0.8}N$ DBR was grown thereon by ammonia molecular beam epitaxy (MBE). The fabrication of the patterned substrates, the growth procedure and the characterization of the DBRs will be detailed elsewhere.[16] If such a DBR, with a total thickness of about 2.5 μm, was grown on plain Si substrate it would display a high density of cracks due to the thermal expansion mismatch problem; however, as can be seen in Figure 1 the same DBR grown on patterned Si is absolutely crack-free, provided the mesa sizes do not exceed 500x500 μm$^2$. This is due to an enhanced strain relaxation enabled by the lateral free surfaces at the mesa edges. Subsequently, 3λ GaN or 7λ/4 ZnO active regions were grown by MBE on top of such DBRs without substrate rotation; this induces a slight thickness gradient (from several nanometers to tens of nanometers across the 2 inch wafer) that allows to finely tune the cavity resonance energy with respect to the exciton one. Finally, the microcavities were completed by the deposition of a top 11 pairs SiO$_2$/HfO$_2$ DBR, as illustrated in Figure 1(b).

Compared to our previous AlN/Al$_{0.2}$Ga$_{0.8}$N-based microcavities, where we had been able to stack only 13 pairs, leading to a modest Q of about 500,[17] the current cavities display Qs in the order of 1500-2500, as shown in Figure 2 (a) and (b), owing to the increased number of crack-free pairs. While a large local Q is an important figure of merit for a microcavity, as it contributes to lowering the polariton lasing threshold, a high spatial homogeneity of the cavity resonance energy is also necessary to get an extended polariton condensate,[18] rather than strongly localized ones as reported in CdTe cavities.[19] With the aim of addressing this issue, the energy position of the LPB mode has been monitored as a function of position within one mesa. The result is illustrated in Figure 2(c), where a 100×100 μm$^2$ area (i.e. about ¼ of the total mesa area) close to a corner of a squared mesa has been analyzed. Two features are noteworthy: first, the energy scale is less than 10 meV, and second, most pixels in the mesa (indeed more than 90%) show an LPB energy lying in between 3.162 eV and 3.166 eV. To give a more quantitative and precise idea of the spatial homogeneity of the cavity, two regions of $10 \times 10$ $\mu m^2$ have been chosen and the LPB energy, in each of their 25 ($2 \times 2$ $\mu m^2$) pixels, has been plotted in the column chart in figure 2 (d). In the "less homogenous" area (represented by the red square) 75% of the pixel values distribute themselves within 2 meV, whereas for a "very homogenous" area (represented by the black square) 85% of the pixel values peak at precisely the same energy of 3.164 eV. While these Q values and large spatial homogeneity constitute already the state-of-the-art for strongly-coupled microcavities operating in the UV, there is still room for improvement given that Q values in excess of 10000 were expected from the numerical simulations.[16]

One advantage of such a low in-plane photonic disorder is that conventional (without using a microscope objective) angle-resolved photoluminescence (PL) measurements can be performed, enabling us to access large angles (>40°) that are otherwise inaccessible because of the reduced numerical apertures available in the UV (typically around 0.4). The result is displayed for the ZnO cavity in Figure 3 (a), below threshold, and demonstrates unambiguously the strong coupling regime, even if the upper polariton branch is not visible in this material system.[20] Indeed, the strong-coupling regime gives rise to a heavier effective mass of the LPB, compared to the simulated bare cavity mode (dashed line) and to the more photonic Bragg polariton modes, which are visible at angles larger than 40°. Furthermore, as the Bragg modes approach the exciton resonance they are seen to anticross too with the excitonic resonances, confirming that they are also in the strong-coupling regime as observed in previous ZnO cavities.[21] Transfer matrix simulations, in which the band to band absorption is "artificially" removed, allows to determine a $\Omega_{Rabi}$ of ~150 ± 20 meV, dependent on the exact measurement position across the 2" sample. We had previously determined $\Omega_{Rabi}$ as a function of cavity thickness in a ZnO microcavity with two dielectric DBRs:[6] for a cavity thickness of 250 nm (~7$\lambda$/4) a $\Omega_{Rabi}$ of 225 meV was extracted. The difference in $\Omega_{Rabi}$ between the two cavities, which display the same thickness, is due to the larger field penetration depth in the epitaxial DBR compared to the dielectric one.

An alternative way to assess the strong-coupling regime consists in measuring the change in the LPB effective mass as a function of the cavity-exciton detuning (i.e. the energy difference between the bare exciton and the bare cavity modes).[22] This is illustrated with the GaN microcavity in figure 3(c), where the LPB dispersion has been imaged for several detunings, going from slightly positive to negative values and leading to excitonic fractions from 56% (leftmost panel) to 11% (rightmost panel). In this case, the strong-coupling manifests itself as an increase of the LPB effective mass as the bare exciton energy is approached. From transfer matrix simulations of all the acquired spectra, a $\Omega_{Rabi}$ of ~70±10 meV can be extracted. Although this value is much smaller than that for ZnO (consistent with an oscillator strength 4 to 6 times smaller for GaN than ZnO), it is the largest to date for any pure planar GaN microcavity. Indeed, the cavity thickness was intentionally made larger than what is common in the *polaritonics* community (3$\lambda$ instead of typically 3$\lambda$/2)[11,23] with the aim of increasing the exciton-cavity mode overlap and enhance the $\Omega_{Rabi}$. The goal here was to increase the depth of the polariton trap in reciprocal space and to reduce as much as possible the thermal escape of polaritons at room temperature.[5]

The nonlinear properties of both cavities were studied under "large spot" excitation conditions. As shown in Figure 4(a) for the ZnO cavity (pumping spot diameter ~ 50 μm), and in Figure 4(b) for the GaN cavity (pumping spot diameter ~ 15 μm), both show a nonlinear intensity rise above a given threshold, which is accompanied by a drastic reduction of the LPB full-width at half maximum and an accumulation of polaritons at the bottom of the LPB (see Figure 3(b)). While a detailed comparison of the complete condensation phase diagrams (i.e. the condensation threshold as a function of temperature and detuning, as well as a function of pumping spot size) will be given elsewhere, the feature that should be highlighted here is the relatively small blueshift below threshold compared to the $\Omega_{Rabi}$; indeed, the blueshifts amount to 3% of the $\Omega_{Rabi}$ in the ZnO microcavity and to 5% of the $\Omega_{Rabi}$ in the GaN one for the detunings studied in Figure 4. Hence, the polariton lasing modes remain far from the bare cavity modes, as indicated by the dashed lines in Figures 4 (a) and (b), proving that lasing occurs in the strong-coupling regime and that no transition towards the weak-coupling regime occurs.[24]

To conclude, we have fabricated on the same common platform GaN and ZnO optical microcavities displaying room temperature polariton lasing. The approach introduced in this letter, namely the patterning of the silicon substrate, allows to combine in one and the same microcavity the advantages of crystalline active materials, obtained by MBE, and of high Qs, which have been enabled by the possibility of stacking a large number of AlN/AlGaN pairs without generating thermal cracks. This platform can be regarded as an ideal building block for future electrically-injected room temperature polariton devices.

This work was partially supported by the EU under contract FP7 ITN Clermont 4 (235114) and by GANEX (ANR-11-LABX-0014). The authors would like to acknowledge D. Solnyshkov and G. Malpuech for fruitful discussions.

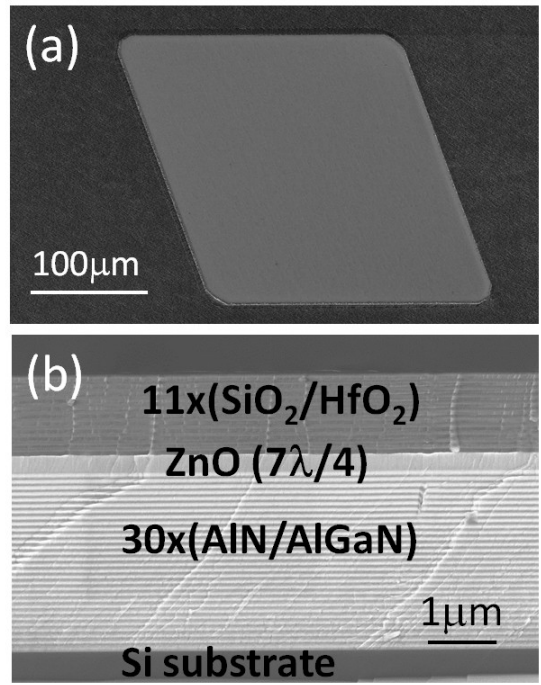

**Figure 1**

FIG.1. Scanning electron microscopy images of the fabricated microcavities. (a) Plain-view image of a diamond-shaped mesa patterned on a Si(110) substrate with a $30\times(AlN/Al_{0.2}Ga_{0.8}N)$ DBR grown thereon that shows no cracks. (b) Cross-section view of the complete ZnO microcavity, which is several micrometers thick. For the GaN microcavity, the structure is essentially the same except that the active region is $3\lambda$ thick.

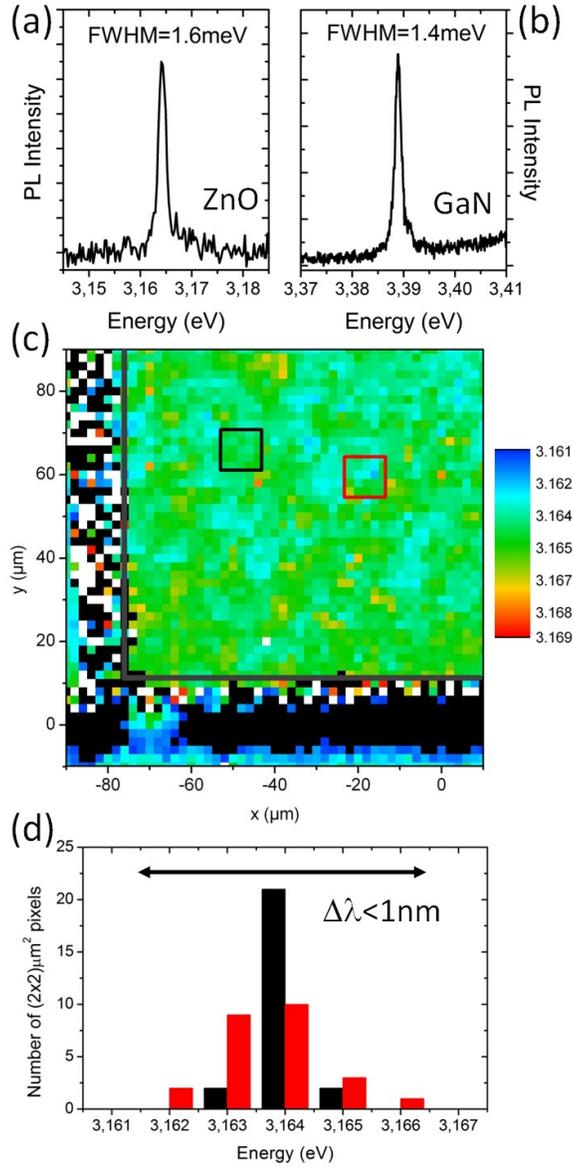

**Figure 2**

FIG.2. Quality factor and spatial homogeneity of the microcavity photonic properties. (a) Typical μPL spectrum, below the lasing threshold, measured at an angle of 0° on the ZnO microcavity. The linewidth of 1.6 meV would correspond to an effective Q of ~1975. (b) Typical μPL spectrum, below the lasing threshold, measured at an angle of 0° on the GaN microcavity. The linewidth of 1.4 meV would correspond to an effective Q of ~2420. (c)

µPL mapping of the LPB energy for the ZnO cavity (scale in eV) at 300K as a function of the position on the mesa close to one of its corners. Each pixel has a size of 2×2 $\mu m^2$ and the excitation spot has a diameter smaller than 1 µm. (d) Distribution of the LPB energy inside the black ("more homogeneous") and red ("less homogeneous") squares depicted in (c). Each square has a 10×10 $\mu m^2$ area and contains 25 pixels. The energy distribution is less than 1 nm wide.

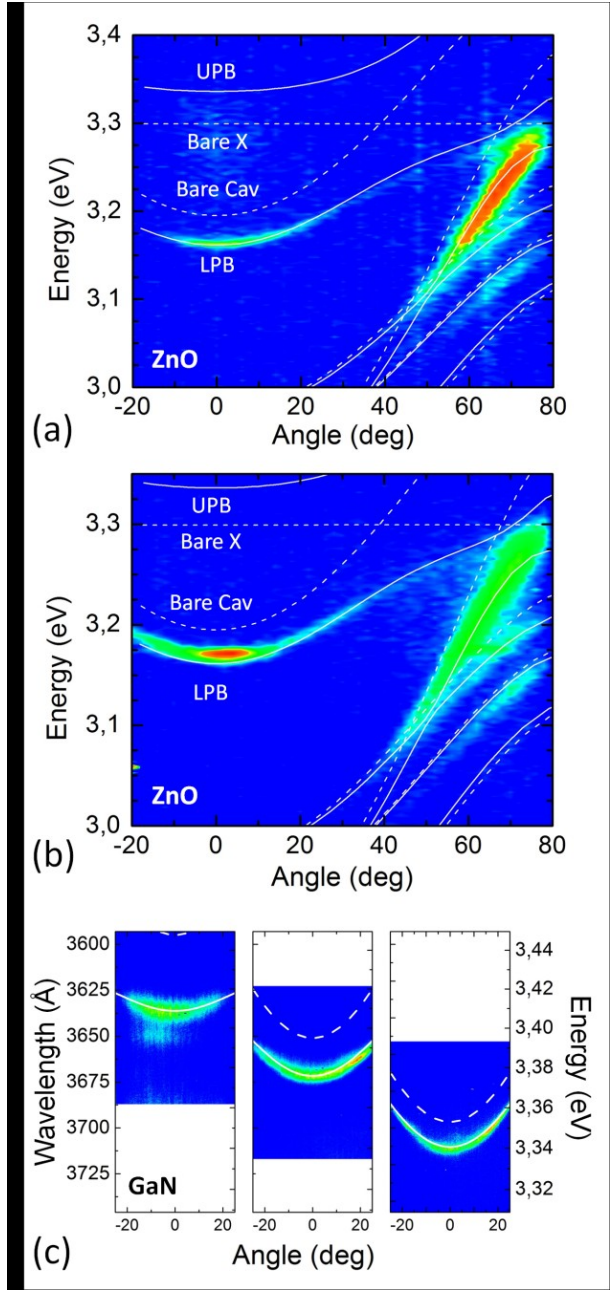

**Figure 3**

FIG.3. Strong-coupling at room temperature. (a) Angle-dependent PL measured on the ZnO

microcavity below threshold with a large excitation spot (>50 μm) for TE polarization. (b)

Angle-dependent PL measured on the ZnO microcavity above threshold ($1.05 \times P_0$) with a large excitation spot ($>50 \ \mu m$) for TE polarization. The simulated bare cavity and Bragg modes (dashed lines) and polariton modes (solid lines) are displayed in (a) and (b), where a logarithmic intensity scale has been used. (c) Fourier space images of the GaN microcavity for different detunings (intensity in linear scale), illustrating the LPB effective mass variation as its excitonic/photonic fraction changes. The detunings and excitonic fractions are: +8 meV and 56% (left panel), -45 meV and 23% (central panel) and -87 meV and 11% (right panel). The simulated bare cavity mode (dashed line) and LPB (solid line) are both shown.

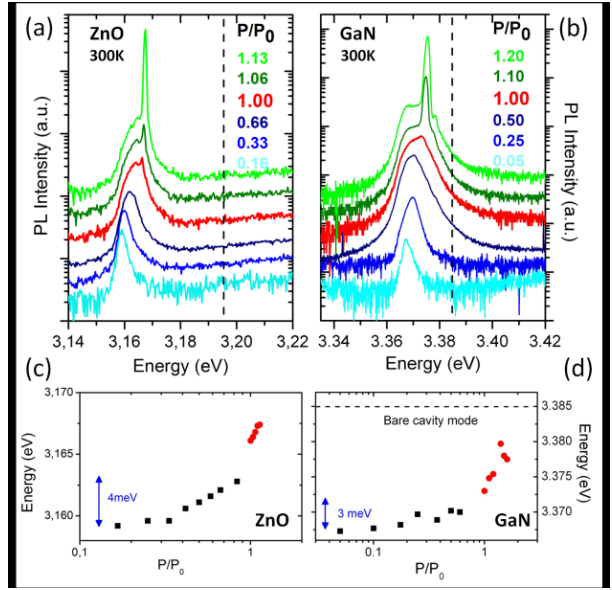

FIG.4. Room temperature polariton lasing. (a) Power dependent series acquired under "large spot" and quasi-continuous excitation conditions (Nd:YAG laser with 4 kHz repetition rate and pulses of 400 ps duration) on the ZnO microcavity at room temperature. The dashed line indicates the position of the bare cavity mode. (b) Power dependent series acquired under "large spot" and pulsed excitation conditions (Ti:Sapphire laser with 76 MHz repetition rate and pulses of 130 fs duration) on the GaN microcavity at room temperature. The dashed line indicates the bare-cavity mode. (c) and (d) Energy of the LPB (black squares) and polariton lasing mode (red circles) as a function of excitation power as extracted from (a) and (b), respectively. Blue arrows show the amount of LPB blueshift below threshold. Note that the vertical scales are not the same in (c) and (d).